\documentclass{aa}
\usepackage{txfonts}
\usepackage{graphicx}
\begin{document}
   \title{Intermediate to low-mass stellar content of Westerlund 1
\thanks{Based on observations collected at the European Southern Observatory,
La Silla, Chile, and retrieved from the ESO archive (Prog ID 67.C-0514).
}
}

   \author{Wolfgang Brandner\inst{1,2} \and
          J.S.\ Clark\inst{3} \and
          Andrea Stolte\inst{2} \and
          Rens Waters\inst{4} \and
          Ignacio Negueruela\inst{5} \and
          Simon P.\ Goodwin\inst{6}}

   \offprints{Wolfgang Brandner}

   \institute{Max-Planck-Institut f\"ur Astronomie, K\"onigstuhl 17,
              69117 Heidelberg, Germany\\
              \email{brandner@mpia.de}
         \and
             UCLA, Division of Astronomy, Los Angeles, CA 90095-1547, USA
         \and
             Open University, UK
         \and
	    Sterrenkundig Instituut ``Anton Pannekoek'', Amsterdam, The Netherlands
         \and
	    Dpto.\ de F\'{\i}sica, Ingenier\'{\i}a de Sistemas y Teor\'{\i}a de la Se\~{n}al, Universidad de Alicante, 03080 Alicante, Spain
	\and
	Dept.\ of Physics \& Astronomy, University of Sheffield, Sheffield, UK
             }

   \date{Received ---; accepted ---}

   \abstract{
We have analysed near-infrared NTT/SofI observations of the starburst cluster 
Westerlund 1, which is among the most massive young clusters in the Milky Way.
A comparison of colour-magnitude diagrams with theoretical main-sequence and 
pre-main sequence evolutionary tracks yields improved extinction and distance
estimates of A$_{\rm Ks}$ =  1.13$\pm$0.03\,mag and d = 3.55$\pm$0.17\,kpc
(DM  = 12.75$\pm$0.10\,mag). The pre-main sequence population is best
fit by a Palla \& Stahler isochrone for an age of 3.2\,Myr, while the main 
sequence population is in agreement with a cluster age of 3 to 5\,Myr.
An analysis of the structural parameters of the cluster yields that the 
half-mass radius of the cluster population increases towards lower mass, 
indicative of the presence of mass segregation. The cluster is clearly 
elongated with an eccentricity of 0.20 for stars with masses between 10 and 
32\,M$_\odot$, and 0.15 for stars with masses in the range 3 to 10\,M$_\odot$.
We derive
the slope of the stellar mass function for stars with masses between 3.4 and
27\,M$_\odot$. In an annulus with radii between 0.75 and 1.5\,pc from the
cluster centre, we obtain a slope of  $\Gamma = -1.3$. 
Closer in, the mass function of Westerlund 1 is shallower with
$\Gamma = -0.6$. The extrapolation of the mass function for stars with masses
from 0.08 to 120\,M$_\odot$ yields an initial total stellar mass of 
$\approx$52,000\,M$_\odot$, and a present-day mass of 20,000 to 45,000\,M$_\odot$
(about 10 times the stellar mass of the Orion Nebula Cluster, and 2 to 4 times 
the mass of the NGC 3603 young cluster), indicating that Westerlund\,1
is the most massive starburst cluster identified to date in the Milky Way.
   }

   \maketitle
%

\section{Introduction}

Clustered star formation is the dominant mode of star formation in
the Universe with about 80\% of all stars originating in giant molecular clouds
 (Lada \& Lada \cite{lada}). The globular cluster systems in
the Milky Way or M\,31 might be remnants of an epoch of Super Star Cluster
(SSC) formation early in the history of these galaxies. Low-mass
stars lost in the course of the dynamical evolution of the SSCs could 
explain the majority of Pop II field stars found in galaxy halos
(Kroupa \& Boily \cite{kroupa02}). While the Galactic globular clusters 
formed more than 12\,Gyr ago, 100s of young SSCs with masses comparable to or greater than Galactic globular clusters; i.e., present-day masses 
$\ge 100,000\,{\rm M}_\odot$, have been identified
in interacting galaxies like, e.g., the \object{Antennae galaxies} 
(Whitmore \& Schweizer \cite{whitmore95}). 

At extragalactic distances, however, individual cluster members are often 
unresolved, and physical properties of SSCs  -- such as their initial mass 
function (IMF) and total mass, degree of binarity, or density profile -- 
must be  estimated from their integrated spectra and photometry alone,  
which in turn are  dominated by the most massive, luminous stars present.
The contribution of lower mass stars to the total cluster mass, which 
is crucial for the dynamical evolution and the long-term survival of a 
cluster (e.g., de Grijs \& Parmentier \cite{degrijs07}, and references
therein), remains uncharted.
Indirect estimates of the total cluster mass can be derived by
extrapolating the IMF towards lower masses. Another possibility 
is to determine the dynamical mass of a cluster from its 
half mass radius and velocity dispersion 
under the assumption of virial equilibrium. Studies by different groups
either agreed with 
a Kroupa (\cite{kroupa02a}) IMF over the whole range of stellar masses
(e.g. Larsen \& Richtler \cite{larsen}) or found a truncated IMF towards lower 
stellar masses
(e.g., Rieke et al.\ \cite{rieke93}; McCrady et al.\ \cite{mccrady05}). 
Young, massive clusters, however, are not necessarily in virial 
equilibrium, as photoionization and winds from O-stars in their centres rapidly
expel any remnant gas left over from the formation process (Kroupa \& Boily 
\cite{kroupa02};  Bastian \& Goodwin \cite{bastian06}).

The determination of the IMF slope as well as the potential presence or 
absence of a low mass stellar component, is critical for constraining 
the long term evolution of such clusters.
This addresses the 
question of whether (proto-) Globular Clusters  are still forming in the 
Universe today. Moreover, as the majority of 
low mass field stars might have formed in clusters, the uncertainty in the 
initial ratio of high to low mass stars in clusters  results 
in a corresponding uncertainty in the star formation and chemical 
enrichment history of galaxies like our Milky Way.

Therefore we examine starburst clusters in the Milky Way and its companion 
galaxies,
where SSCs can be resolved into individual stars, and cluster properties like the IMF can be determined directly. 
As a {\it starburst} cluster we consider any cluster which is massive
enough to house several early O-type stars with initial masses in excess
of 50\,M$_\odot$ in its centre. 
Among the most massive Galactic starburst clusters studied to date
are \object{NGC\,3603YC} (e.g., Brandl et al.\ \cite{brandl}, Stolte et al.\ \cite{stolte06}) and \object{Westerlund 2} (Wd 2,
Ascenso et al.\ \cite{ascenso07}),
both in the Carina spiral arm,  and the two massive clusters \object{Arches} 
(Stolte et al.\ \cite{stolte02,stolte05}; Figer et al. \cite{figer02}; Kim et al.\ \cite{kim06}) and
\object{Quintuplet} (e.g., Cotera et al.\ \cite{cotera96}, Figer et al.\ \cite{figer04})   in the Galactic centre (GC) region.

One questions of particular interest is whether or not the presence
of high mass stars hinders or even quenches the formation of low mass stars.
Intense UV radiation fields and strong stellar
winds of nearby OB stars might lead to the destruction  of
protostellar cores before low-mass stars can form.
Furthermore, in extreme environments like the GC region,
variations in the physical conditions leading to higher thermal and/or 
turbulent  pressure, as well as the presence of strong magnetic fields 
might preferentially yield higher mass prestellar cores than in more benign
environs (Morris \cite{morris93}, Stolte et al. \cite{stolte05}, Larson \cite{larson06}, Klessen et al.\ \cite{klessen07}).

Along with the Arches and Quintuplet clusters in the GC region,
and the NGC 3603YC in the Sagittarius-Carina spiral arm, 
\object{Westerlund 1} (Wd 1), located in the Scutum-Crux spiral arm, is one of the very
few young, massive starburst clusters in the Milky Way. 

Discovered by Westerlund (\cite{west61}), subsequent optical observations 
of the galactic open cluster  Westerlund 1 (Wd~1) suggested an unusually 
rich population of cool and hot  supergiants (Westerlund \cite{west87}), 
from which a large cluster mass could be inferred. Recent optical
spectroscopic and photometric observations have confirmed these 
assertions (Clark \& Negueruela \cite{cn}; Clark et al. \cite{clark05} 
(henceforth C05); Negueruela \& Clark \cite{nc}). 
Located close to the Galactic plane (b = $-0.35^\circ$) at a distance of $\sim$3--5\,kpc, Wd\,1 is subject to significant foreground extinction. Thus optical studies of the cluster were limited to the most massive stars  of Wd1.
C05 spectroscopically identified about 50 cluster members with masses in excess of 30\,M$_\odot$. Adding 150 probable cluster members identified by means of photometry, they estimate that the high-mass stellar content of Wd\,1 alone amounts to 6000\,M$_\odot$.
Assuming a Kroupa IMF to extrapolate from this population, C05 then infer a likely total mass of $\sim 10^5$\,M$_{\odot}$  for Wd1; directly comparable  to the 
masses of SSCs observed in other galaxies.  

Up to now, however, only the evolved, high-mass stellar population of the cluster has been characterised. The presence of Wolf-Rayet stars (e.g., Crowther et al.\ \cite{crowther06}; Skinner et al.\ \cite{skinner06a}), a pulsar indicative of a recent supernova (Skinner et al.\ \cite{skinner06b}, Muno et al.\ \cite{muno06a}), as well as O supergiants,
suggest an age between 3 and 5\,Myr for the cluster. Because of a visual foreground extinction of A$_{\rm V} \approx 10$\,mag, and the uncertainties in the intrinsic luminosity of the evolved, massive stars, distance determinations to Wd\,1 range from 2 to 5.5\,kpc (C05).

Because of its proximity, Wd1 provides the unique opportunity to directly 
determine the spatially resolved IMF of a SSC and, 
given that it is located well away  from the GC in a 
supposedly `quiescent' region of the Galactic disc, potentially study
physical properties imprinted by the formation process from the parental
molecular cloud on the cluster. 
With the aim to better determine basic cluster properties such as age, 
foreground extinction and distance, to directly measure the IMF, and to study
the structural properties of Wd\,1, we have analysed deep near IR 
observations. This paper describes the 
analysis of  NTT/SofI observations of Wd1 covering a $\sim 5' \times 5'$ 
field of view centred on the cluster core, which  are sensitive 
to $\sim$ solar mass stars at a distance of 4\,kpc; a second  paper 
details subsequent VLT/NAOS-CONICA adaptive optics observations of 
selected core regions sensitive to $\sim$0.2\,M$_{\odot}$.

The structure of the paper is as follows: in section 2 we present the 
observations and the data analysis. The basic astrophysical quantities
extinction, distance, and age of Wd~1 are derived in section 3.
Section 4 discusses the mass function and the total stellar mass of Wd~1.
Dynamical and structural parameters of Wd~1 are addressed in section 5, followed by a comparison
of Wd~1 to other well studied starburst clusters in section 6. Section 7
summarises the results.

\begin{figure*}[htb]
\centerline{
\hbox{
 \includegraphics[width=8.5cm,angle=0]{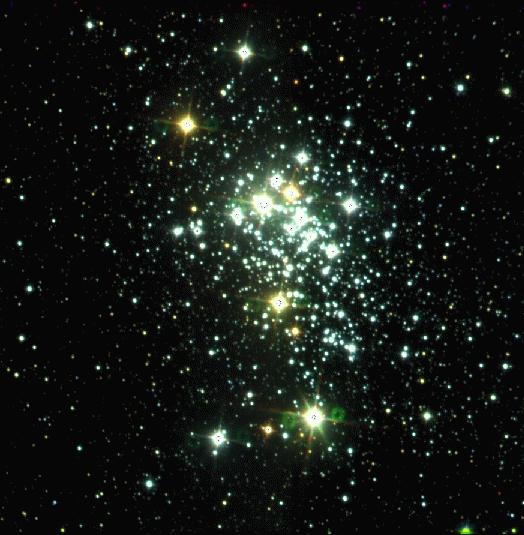}
 \includegraphics[width=8.6cm,angle=0]{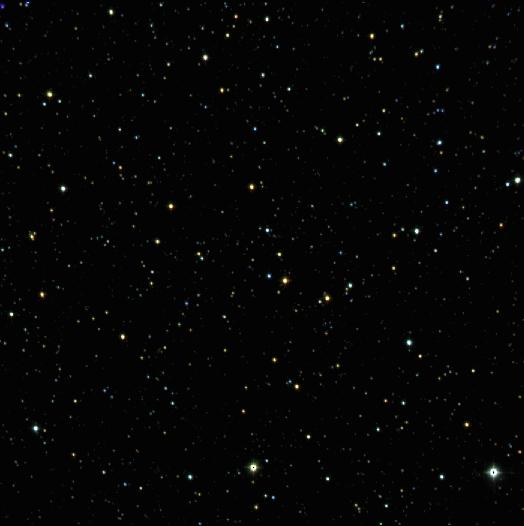}
}
}
\caption{Composite colour JHKs images of Westerlund 1 (left)
and the comparison field (right) obtained with NTT/SofI. The field of
view is $4' \times 4'$. North is up, and east is to the left.}
\label{image_rgb}
    \end{figure*}

\section{Observations and data reduction}

\subsection{NTT/SofI observation}

JHKs-broad band imaging observations of Wd 1 (centred on
RA(2000) = 16$^h$47$^m$03$^s$, DEC(2000) = -45$^\circ$50$'$37$''$)
and a nearby comparison field (located $\approx 7'$ to the east, and
$\approx 13'$ to the south of Wd 1, and centred on RA(2000) = 
16$^h$47$^m$43$^s$, DEC(2000) = -46$^\circ$03$'$47$''$), each covering an area 
of $4.5' \times 4.5'$
(corresponding to $4.8\,{\rm pc} \times 4.8\,{\rm pc}$
at the distance of Wd 1), obtained with NTT/SofI (PI: J. Alves) were retrieved from the ESO archive (see Fig.\ \ref{image_rgb}).
SofI is equipped with a Hawaii HgCdTe detector. The observations were
obtained with a plate scale of 0.29$''$/pixel.
Individual integration times were 1.2s (DIT) and 10 frames were co-added,
resulting in 12s of integration time per dither position. For each field and
filter a total of 10 dither positions were used, resulting in a
total integration time of 120s in each filter.

\subsection{Eclipse raw data reduction}

Data reduction was carried out
using the eclipse jitter routines (Devillard \cite{devil01}). Because of the
high degree of crowding in the cluster field, sky frames derived from the
comparison field were used in the reduction of the Wd\,1 frames.
The resolution on the final images is $\approx 0.75''$ to $0.80''$ (see
Table \ref{obslog} for more details).

\begin{table}[htb]
\caption[]{Observing log for the NIR data. The observations were
carried out with the ESO NTT and SofI on 9 June 2001.}
         \label{obslog}
\begin{tabular}{lcccr}
 Object &  Filter      & t$_{\rm int}$     &  seeing        & sat.\ limit  \\
        &              & [s]               &   [$''$] & [mag]  \\
            \hline
            \noalign{\smallskip}
Westerlund 1   & J  & 120 & 0.81 & 10.1 \\
Westerlund 1   & H  & 120 & 0.82 & 9.5 \\
Westerlund 1   & Ks & 120 & 0.73 & 9.1 \\
Off-Field & J  & 120 & 0.77 & 10.2\\
Off-Field & H  & 120 & 0.77 & 9.6 \\
Off-Field & Ks & 120 & 0.74 & 9.1\\
            \noalign{\smallskip}
            \hline
\end{tabular}
   \end{table}

\begin{figure*}[htb]
   \centering
 \includegraphics[width=12.5cm,angle=90]{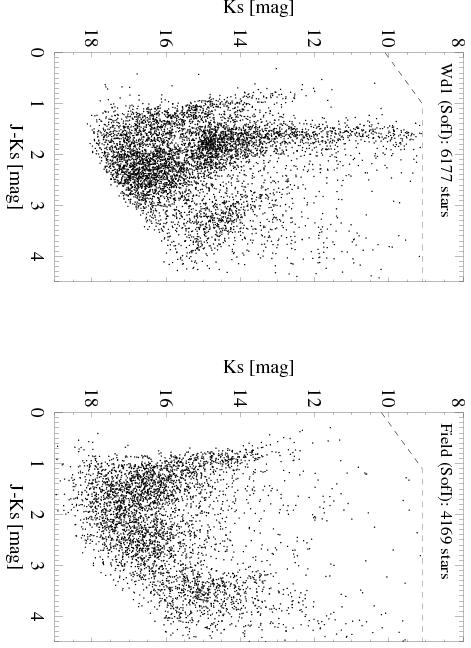}
\caption{Colour magnitude diagram of the cluster (including field
contamination, left), and the comparison field (right). The dashed
line marks the saturation limit.}
\label{cmd_all}
    \end{figure*}

\subsection{Photometric analysis and calibration}

PSF fitting photometry in J, H, and Ks for 7000 stars in the Wd1 area, and 
5300 stars in the off-field was derived using the IRAF implementation of 
DAOPHOT (Stetson \cite{stetson87}). Photometric zeropoints and colour terms were computed by comparison
of instrumental magnitudes of relatively isolated, bright sources with 
photometry from the 2MASS point source catalogue. 

The resulting Ks vs.\ J--Ks colour-magnitude diagrams (CMD) for Wd1
and the off-field for all stars with DAOPHOT fitting errors $\le$0.2\,mag
are presented in Fig.\ \ref{cmd_all}. The
dashed line indicates the saturation limit (see Table \ref{obslog}).
The off-field is characterised by a blue main-sequence of foreground
stars, and a red sequence of background stars, most of which could be
red giants located in the galactic bulge. The Wd~1 field shows the
main-sequence and pre-main sequence members of the cluster in addition 
to the 
two populations visible in the off-field. Note that the red background
population appears to be less reddened close to Wd~1 compared to the off-field.
This is most likely due to varying extinction along the line of sight towards
the bulge at distances larger than the distance to Wd~1.

In the following analysis, only stars with DAOPHOT photometric fitting
errors of 0.2\,mag or less are considered.

\begin{figure*}[htb]
\centerline{
\hbox{
 \includegraphics[width=8.0cm,angle=0]{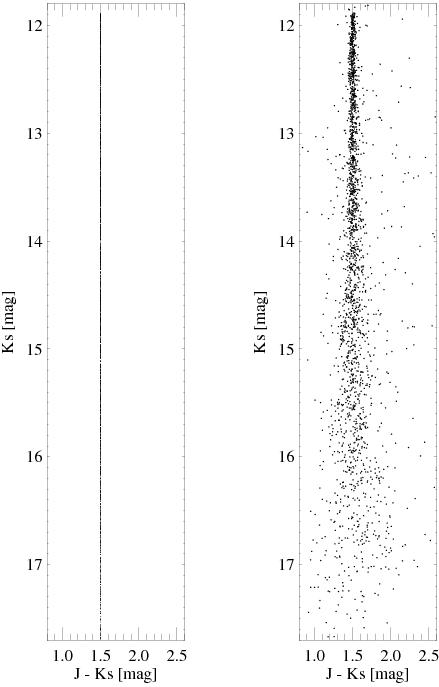}
 \includegraphics[width=8.3cm,angle=0]{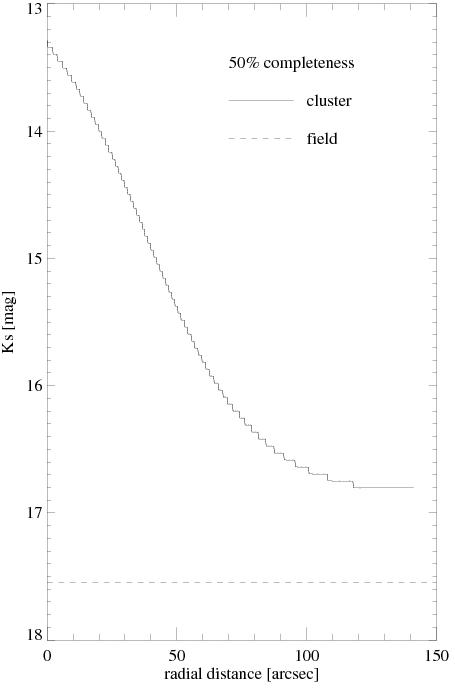}
}
}
\caption{Results of the completeness simulations for Westerlund 1. The two figures on the left show the input magnitudes and colours and the recovered magnitudes and colours, respectively, for the cluster frame.
From this we also derive the photometric uncertainties.
The figure on the right shows the  50\% completeness limits. For the cluster, increasing crowding towards its center pushes the completeness limit (solid line) towards brighter magnitudes. The homogeneous stellar density in the comparison field results in an overall  50\% completeness level down to Ks = 17.55\,mag (dashed line).}
\label{inc_sim}
\end{figure*}

\subsection{Incompleteness characterisation}

Incompleteness simulations with the aim to assess the effect of crowding on
our ability to detect faint sources were carried out both for the jitter
combined off-field and the cluster field. In order not to change the crowding 
characteristics of the frames, for each run only 50 stars were added using 
{\sl addstars} under DAOPHOT. Artificial star magnitudes were allowed to scatter $\pm$0.5\,mag
around a predefined value, and the stars were placed at random positions
on the Ks-band frames. The detection experiment was following the same steps 
and using the same PSF as the initial DAOPHOT analysis. Only those stars,
whose recovered magnitudes match to within $\pm$0.5\,mag the input magnitude,
were counted in the final analysis. For each predefined
magnitude, 10 such frames were created and analysed, i.e., 10$\times$50 stars
= 500 stars/mag bin. Then the 
magnitude was increased by 1.0\,mag, and 10 new frames were created and
analysed. This procedure was repeated $6 \times$, probing the magnitude range Ks = 11.9\,mag to 17.9\,mag for the cluster frame, and Ks = 13.9 to 19.9\, mag
for the field frame. For the
J-band incompleteness simulations, stars were placed at the same positions
as on the Ks-band frames. J-band magnitudes were computed assuming J--Ks = 
2.0\,mag, which is an intermediate colour between the (lower) main-sequence 
stars in Westerlund 1 with J--Ks $\approx$ 1.5\,mag, and the pre-main sequence
stars with J--Ks $\approx$ 2.3\,mag.
 
In total 240 frames with artificial stars added were analysed. The results
are summarised in Fig.\ \ref{inc_sim}. Note that the lack of bright stars
in the off-field pushes the 50\% completeness limit to fainter magnitudes
compared to the cluster field.

\subsection{Statistical field subtraction}

With Galactic coordinates l~$\approx 339.55^\circ$ and b~$\approx -0.40^\circ$,
Wd~1 is embedded in a rich population of fore- and background stars.
The more than 4000 stars detected in J and Ks in the comparison field with
DAOPHOT fitting errors $\le 0.2$\,mag
(Fig.\ \ref{cmd_all}, right) give a good estimate of the field star population
mix. This provides a firm basis to remove the field star contamination in the 
cluster CMD, and hence determine a clean, relatively unbiased cluster population for further analysis.

\begin{figure*}[htb]
   \centering
\hbox{
 \includegraphics[width=5.8cm,angle=0]{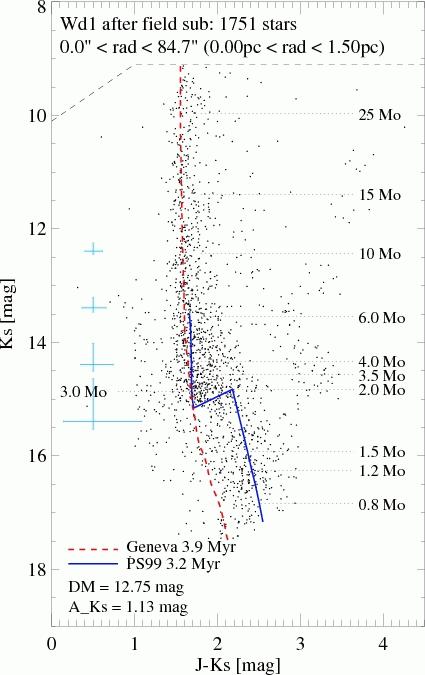}
 \includegraphics[width=5.8cm,angle=0]{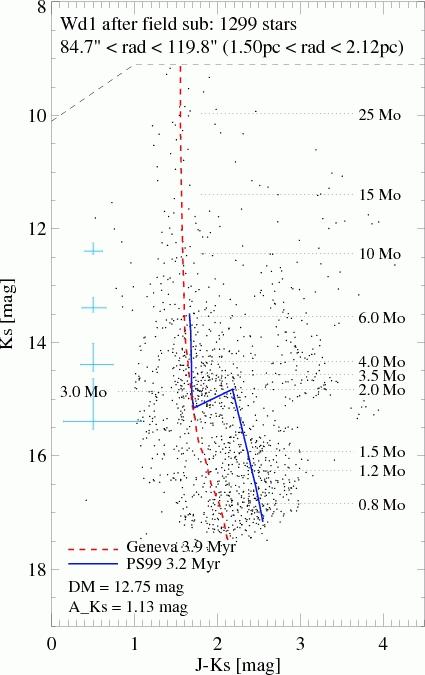}
 \includegraphics[width=5.8cm,angle=0]{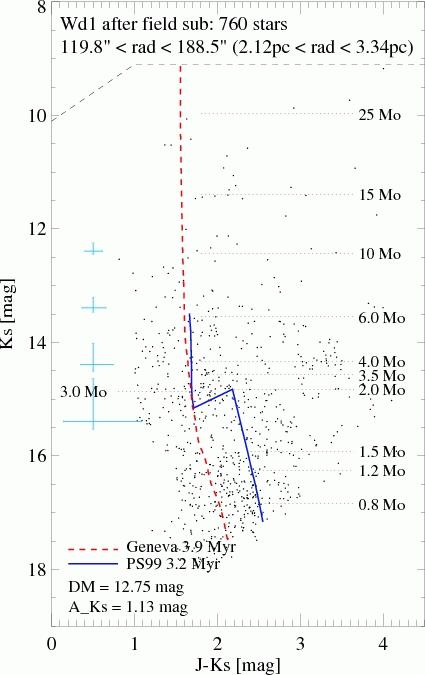}
}
\caption{Field subtracted CMDs of the cluster for 3 annuli. A 3.9 Myr Geneva
(Z=0.020, with overshooting and OPAL opacities)
and a 3.2 Myr Palla \& Stahler isochrone are overplotted for comparison.
The dashed line marks the saturation limit. Photometric uncertainties in K-magnitude and J--K colour as determined by the artificial star tests to determine incompleteness are plotted to the left of the cluster sequence. The dotted lines indicate stellar masses of cluster members according to the Geneva isochrone for masses $\ge 6.0$\,M$_\odot$, and the Palla \& Stahler isochrone for masses less than 6.0\,M$_\odot$.}
\label{cmd_iso}
    \end{figure*}

The statistical field subtraction is based on a comparison of the cluster
and the field CMD. The CMDs are subdivided into grid cells
with a step size of 0.5\,mag in colour and magnitude. The number of
field stars within each cell is counted,
normalised to the ratio of the sky areas covered by the image and the
selected cluster annuli, and corrected for the differences in completeness
fraction between the field and the cluster (see Fig.\ \ref{inc_sim}). Finally, the same number of
stars is subtracted at random from the corresponding grid cell in the
cluster CMD. Examples of the cleaned-up (``field subtracted'') cluster
CMDs are shown in Fig.\ \ref{cmd_iso}.

\begin{figure}[htb]
   \centering
 \includegraphics[width=7.5cm,angle=0]{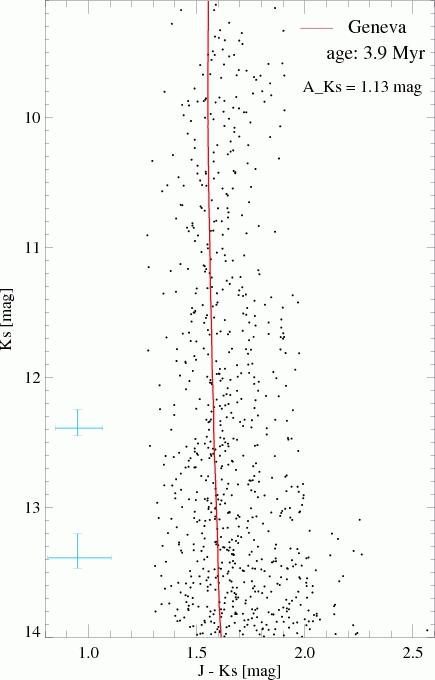}
\caption{Colour magnitude diagram of the main sequence population of the cluster
for stars with DAOPHOT fitting uncertainties smaller than 0.05\,mag in J and Ks.
A Geneva isochrone (solid line, age of 3.9 Myr, DM = 12.75\,mag) for  A$_{\rm K} = 1.13$\,mag is overplotted. Photometric uncertainties (root mean square) as determined by the artificial star test are indicated to the left of the cluster 
main sequence. Even though some of the stars with colours 0.3 to 0.5\,mag 
redder than the main-sequence have been identified as W-R stars, the relatively
wide colour range is in part due to differential foreground reddening.}
\label{cmd_upperms}
    \end{figure}

\section{Extinction, cluster age, distance and evolutionary phases of cluster members}
A starburst cluster most likely forms in a single event (``burst''), resulting in a small age spread of its stellar population.
Galactic starburst clusters like NGC\,3603 or Arches with several 1,000 to 
10,000 stars spanning a factor of 100 or more in individual stellar mass have the advantage that 
global properties like foreground extinction, distance or age can be derived 
with a relatively small uncertainty by comparison to theoretical isochrones.
For a detailed analysis of the mass function and the dynamical properties
of Wd\,1, these basic astrophysical quantities have to be derived first. 

For the following analysis we use Geneva isochrones (Lejeune \& Schaerer 
\cite{lejeune01}) describing main-sequence and
post-main sequence evolution of stars with masses between 0.8 and
120\,M$_\odot$, and Palla \& Stahler isochrones (Palla \& Stahler 
\cite{palla99}) describing pre-main sequence 
and main-sequence evolution of stars with masses between 0.1 and 6\,M$_\odot$.
In the following paper describing the analysis of the adaptive optics data,
we will provide a more in-depth comparison of evolutionary tracks.
Isochrones by Palla \& Stahler are preferred over other, more
recent tracks (e.g., Siess et al.\ \cite{siess00}) as they provide a better
fit to the pre-main sequence / main sequence transition region in
near infrared CMDs of starburst clusters. In their comparison of NIR
observations of NGC\,3603\,YC with PS99 and Yonsei-Yale 
(Yi et al.\ \cite{yi01}) isochrones, Stolte et al.\ (\cite{stolte04})
also found a better fit of the PS99 tracks to the observed 
pre-main sequence / main sequence transition region.
Starburst clusters, i.e.\ clusters housing early O-type stars in their centres,
 might be special in this respect, as the intense UV radiation of the most massive cluster members rapidly photo-evaporates any remaining accretion disks or envelopes around lower mass stars in the cluster (e.g., Brandner et al.\ \cite{brandner00} and references therein). Thus pre-main sequence evolution in starburst clusters seems to be well described by non-accreting tracks.
  For a thorough discussion of  short-comings of theoretical isochrones compared to observations we refer to Mayne et al.\ (\cite{mayne07}). 
 Using the same set of isochrones as previously used for the analysis of the
mass functions of NGC 3603 and Arches also facilitates a direct comparison
of the 3 clusters. We use a mass of 6.0\,M$_\odot$ as the transition point from Geneva to Palla \& Stahler isochrones as both sets of tracks predict the same NIR magnitudes for stars of this mass in the age range considered.

\begin{table}[htb]
\caption[]{Basic astrophysical parameters extinction, DM (distance), 
metallicity and age
of Wd\,1 as derived from comparison of the Ks vs.\ J--Ks CMD with theoretical
Geneva and Palla \& Stahler isochrones.}
         \label{basic}
\begin{tabular}{cccc}
A$_{\rm Ks}$  & DM           & Z            & age          \\ 
 mag        & [mag]        &              & Myr          \\ \hline
1.13$\pm$0.03  & 12.75$\pm$0.10   &0.02   &3.6$\pm$0.7  \\
& (3.55$\pm$0.17\,kpc) & & \\
\end{tabular}
\end{table}

The values for extinction, distance modulus and age of the cluster, which we 
present in the following, were derived in an iterative process by fitting isochrones to the SofI and NACO near-infrared data.

\subsection{Extinction}

As the intrinsic J--Ks colours of main sequence stars with masses in the
range 6 to 30\,M$_\odot$ just vary between
$-0.1$ and $-0.2$\,mag, the foreground extinction
can be derived by simply fitting a zero-age main-sequence (ZAMS) 
to the main-sequence
population of Wd\,1and assuming a standard Rieke \& Lebofsky 
(\cite{rieke85}) extinction law. 
In Fig.\ \ref{cmd_upperms} only stars in the SofI data set with DAOPHOT fitting errors less equal 0.05\,mag are shown. The artificial 
star tests carried out as part of the incompleteness simulation indicate that
the DAOPHOT fitting errors underestimate the true photometric errors. 
Interestingly, the recovered magnitudes are on average brighter than the input
magnitudes of the artificial stars. This asymmetry in the photometric errors
around zero is also present in the colour estimate J-Ks, though less pronounced. 
The photometric errors explain part of the observed scatter in the colour of the main sequence stars. By comparing CMDs for different regions in our field 
of view, we also see evidence for differential extinction. In general, the 
regions to the west and south of the cluster centre suffer slightly lower
foreground extinction than the regions to the east and north of the cluster 
centre.
As a best fit, we get A$_{\rm Ks}$ = 1.13$\pm$0.03\,mag.
Since all our measurements are in the near-infrared, and the theoretical isochrones used have been transformed to JHK-magnitudes and colours, the results are relatively insensitive to any deviation from a standard extinction law. 

In their analysis of the NIR colours of 18 W-R stars in Wd\,1 Crowther
et al.\ (\cite{crowther06}) derived A$_{\rm Ks} = 0.96 \pm 0.14$\,mag, which
within the quoted uncertainties overlaps with the Ks-band extinction derived
by us. We note, however, that for the two W-R stars in overlap with our 
sample (see Table \ref{stars_spt_mass}), Crowther et al.\ (\cite{crowther06}) observe
J-Ks colours 0.35 to 0.46\,mag bluer than we do. Variability as reported
by Bonanos (\cite{bonanos07}) might explain part of this discrepancy, though
there might also be a systematic offset in the zeropoint calibrations
of the two data sets.

For comparison to other extinction determinations, which were exclusively based on studies in the visual, we also compute the visual extinction. Assuming a normal extinction law with R = A$_{\rm V}$ / E$_{\rm B-V}$ = 3.1, A$_{\rm Ks}$ = 1.13\,mag corresponds to A$_{\rm V} \approx 10.1$\,mag (Rieke \& Lebofsky \cite{rieke85}).

Westerlund (\cite{west61}) derived an optical colour V--I  = 
$+4.5$\,mag for the $\approx$80 brightest
stars of the cluster. By assuming an intrinsic colour of (V--I)$_0 = -0.35$,
which is typical for early type stars, this corresponds to E(V-I) = 4.85\,mag.
If one further assumes a standard extinction law with
A$_{\rm V}$ = 1.93 E(V-I) (see Rieke \& Lebofsky \cite{rieke85}),
this results in A$_{\rm V}$ $\approx$ 9.4\,mag. Note that
Westerlund (\cite{west61}) deduced a higher 
A$_{\rm V}$ of 11.2 -- 12\,mag by assuming a ratio between total and selective 
absorption of 2.3$\pm$1.4 (Kron \& Mayall \cite{kron60}). Follow-up 
photometric and spectroscopic observations considering about 260 cluster 
members yielded A$_{\rm V}$ $\approx$ 9.7$\pm$0.8\,mag (Westerlund
\cite{west87}). Clark et al.\ (\cite{clark05}) 
estimate A$_{\rm V}$ = 11.6 to 13.6\,mag, and present evidence
for anomalous extinction towards $\approx$20 OB supergiants in Wd~1.
While for normal interstellar extinction the ratio of total to selective extinction ${\rm R = 3.1}$, R values in the range 1.7 to 5.5 have been observed in high-mass star forming regions (Patriarchi et al.\ \cite{patri01}, but see also Hadfield \& Crowther \cite{hadfield06} and Lamzin \cite{lamzin06}). For R = 3.7, A$_{\rm K} = 1.13 \pm 0.03$\,mag corresponds to A$_{\rm V} \approx 12.0$\,mag, i.e.\ in the range of the visual extinction values determined by Clark et al.\  (\cite{clark05}) towards some of the evolved stars in Wd\,1.

\subsection{Age}

For the age determination, we take advantage of the fact that the width and the shape of the 
pre-main sequence -- main-sequence transition region in J--Ks varies with 
age. This allows us to derive the age of the cluster by comparison with theoretical isochrones.
The best fitting Palla \& Stahler isochrone are for ages of 3.2 to 3.4\,Myr.
Isochrones for an age of 3.0\,Myr or  3.6\,Myr, respectively do not fit as well, in particular when taking the adaptive optics data into account. A more detailed comparison of different sets of theoretical isochrones will be presented in the second paper, which features a detailed analysis of the cluster population in the range of 0.4 to 3\,M$_\odot$. We estimate the age of the low-mass stellar population of Wd\,1 since crossing the ``birthline'' (see Stahler \cite{stahler83}) to 3.3$\pm$0.2\,Myr.

At young ages of $\le$10\,Myr, pre-main sequence evolutionary tracks originating at the stellar birthline, and main-sequence / post-main-sequence evolutionary tracks originating at the zero-age-main-sequence are not necessarily on the same absolute timescale, and might show systematic offsets. This is exemplified by the \object{Orion Nebula Cluster} (ONC), for which Meynet et al. (\cite{meynet93}) fit the youngest Geneva isochrone in their grid for an age of 4\,Myr to the upper main sequence of ONC, while the canonical age for the low mass stars is $<$2\,Myr (Hillenbrand \cite{hillenbrand97}; Palla \& Stahler \cite{palla99}).

In the mass and wavelength (colour) range accessible with the SofI near-infrared data, Geneva isochrones (Lejeune \& Schaerer \cite{lejeune01}) are rather insensitive to age, in particular as near-infrared colours like J--Ks always sample the Rayleigh-Jeans tail in the spectra of the hot, massive stars. Consequently, we cannot use the Geneva isochrones to derive further constraints on the age of Wd\,1. 
Based on the observations of evolved stars in Wd\,1, both Clark et al. (\cite{clark05}) and Crowther et al.\ (\cite{crowther06}) argue for an age 
of 4 to 5\,Myr. In the following analysis, we use Geneva
isochrones in the age range 3.0 to 5.0\,Myr. The isochrones are based on the ``classical'' evolutionary tracks including overshoot, and have been computed for a metallicity of Z=0.02.

As an average age for the low ($3.3 \pm 0.2$\,Myr) and high-mass 
($3.9 \pm 1.0$\,Myr) stellar population of Wd\,1, we adopt 
t$_{\rm cluster}$ = $3.6 \pm 0.7$\,Myr. This also results in consistent
values for the brightness and colours predicted by both sets of tracks for the
6\,M$_\odot$ stars.

\subsection{Distance}

Once extinction and age have been established, the distance to
the cluster follows straight from the observed near-infrared magnitude of the transition
region between pre-main sequence and main-sequence. For the 3.2\,Myr PS99
isochrone, we get DM = 12.75$\pm$0.10\,mag or d = $3.55\pm0.17$\,kpc.

This is in agreement with the study by Westerlund (\cite{west87}),
who derived a distance modulus of DM = 13.6$\pm$0.7\,mag, as well as Clark et 
al.\ (\cite{clark05}), who argue for a distance between 2 and 5.5\,kpc, 
and Crowther et al.\ (\cite{crowther06}), who derive  DM = 13.50$\pm$0.66\,mag
based on the analysis of the NIR colours of 18 W-R stars in Wd\,1. 
More recently, Kothes \& Dougherty (\cite{kothes07}) derived a distance of
$3.9\pm0.7$\,kpc based on the radial velocity of an HI feature likely associated
with Wd\,1, assuming a standard Galactic rotation curve and a distance of 
7.6\,kpc (Eisenhauer et al.\ \cite{eisenhauer05}, Reid \cite{reid93})
between the Sun and the GC.

The results on extinction, age and distance are summarised in Table \ref{basic}.

\subsection{Evolutionary phase of cluster members}

Wd~1 provides a unique, and quite interesting snapshot of stellar evolution
as a function of mass. At an age of $\approx$3.6\,Myr, the most massive cluster
members have already left the main-sequence. Only stars with Ks $\ge 11$\,mag,
corresponding to an MK-type of B0.5V (m $\approx 15$\,M$_\odot$), are still 
close to the ZAMS, whereas more massive stars already started to evolve off
the ZAMS. The cluster main-sequence extends down to Ks $\approx$ 15\,mag,
corresponding to an MK-type of A0V (m = 3\,M$_\odot$). Stars with masses
between 2 and 3\,M$_\odot$ are located in the pre-main sequence -- 
main-sequence transition region, and stars with masses less than 2\,M$_\odot$
are still in the pre-main sequence phase.

Massive young clusters like Wd~1 offer the rare opportunity to bring evolutionary tracks for higher mass stars, starting with the zero-age main sequence, and pre-main sequence evolutionary tracks starting, e.g., with the stellar birthline (Stahler \cite{stahler83}) on the same evolutionary time scale. This is a crucial empirical calibration point in developing a consistent theory of stellar evolution encompassing all stars from 0.08\,M$_\odot$ to 120\,M$_\odot$.

\section{Mass function and cluster mass}

We compute the centre of Wd\,1 from the field-subtracted spatial 
distribution of cluster members. We find the cluster centre to be located
at RA(2000) = $16^{\rm h} 47^{\rm m} 04.0^{\rm s}$, 
DEC(2000) = $-45^\circ 51' 04.9''$.
This location is used as a reference for the following discussion.

\begin{table}[htb]
\caption[]{Mass function slope $\Gamma$ for stars with masses between 3.4 and 27\,M$_\odot$.}
         \label{imf_age}
\begin{tabular}{c|ccc|c}
Radial Dist.\ & 3.2\,Myr & 3.9\,Myr & 4.9\,Myr & 3.9\,Myr   \\
              & z=0.02 & z=0.02 & z=0.02 & z=0.04  \\ \hline
0 -- 0.75\,pc     & $-0.7$ & $-0.6$  & $-0.5$ & $-0.6$  \\
0.75 -- 1.5\,pc   & $-1.4$ & $-1.3$  & $-1.3$ & $-1.4$  \\
1.5 -- 2.1\,pc    & $-1.7$ & $-1.6$  & $-1.5$ & $-1.6$ \\
2.1 -- 3.3\,pc    & $-1.7$ & $-1.7$  & $-1.9$ & $-2.0$ \\
\end{tabular}

{\small
 $\Gamma$ values are quoted for different annuli, and three Geneva isochrones for z=0.020 and for ages of 3.2, 3.9, and 4.9\,Myr, respectively. The column to the right lists $\Gamma$ based on a Geneva isochrone with z=0.040 and for an age of 3.9\,Myr. There is a slight dependence of $\Gamma$ on the age and the metallicity of the adopted isochrone.}
\end{table}

\begin{figure*}[htb]
   \centering
\vbox{
\hbox{
 \includegraphics[width=8.5cm,angle=0]{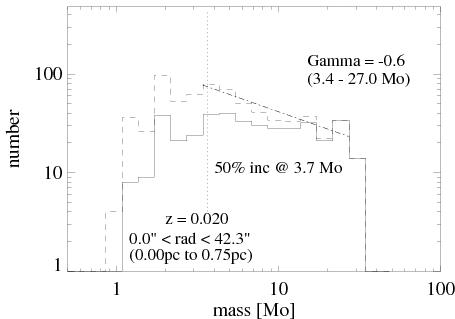}
 \includegraphics[width=8.5cm,angle=0]{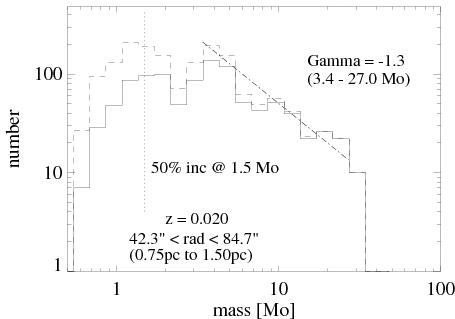}
}
\hbox{
 \includegraphics[width=8.5cm,angle=0]{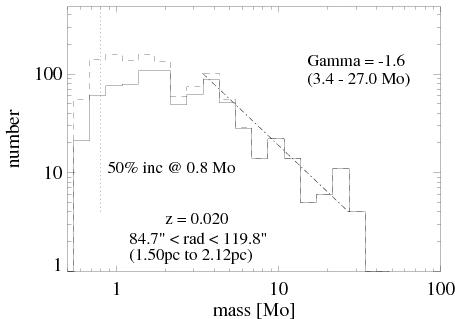}
 \includegraphics[width=8.5cm,angle=0]{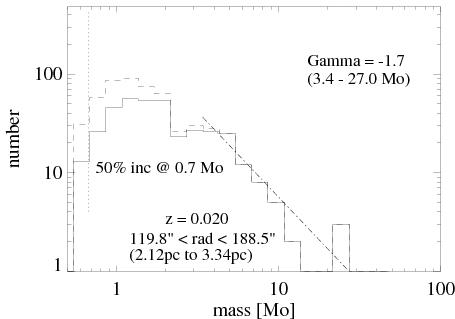}
}
}
\caption{Mass functions (solid line) of the cluster for 4 annuli.
The dashed line gives the incompleteness corrected mass function. The 50\% 
completeness limit is indicated by a vertical dotted line. The mass determination
is based on a 3.9 Myr Geneva isochrone with Z=0.020 for stars with masses $\ge$6.0\,M$_\odot$, and on a 3.2 Myr Palla \& Stahler isochrone with Z=0.019 for stars with mass $<$6.0\,M$_\odot$. The slope of the mass function (dash-dotted line) is computed from stars with masses between 3.4 and 27\,M$_\odot$.}
\label{imf_iso}
    \end{figure*}

\begin{figure}[htb]
   \centering
 \includegraphics[width=6.0cm,angle=90]{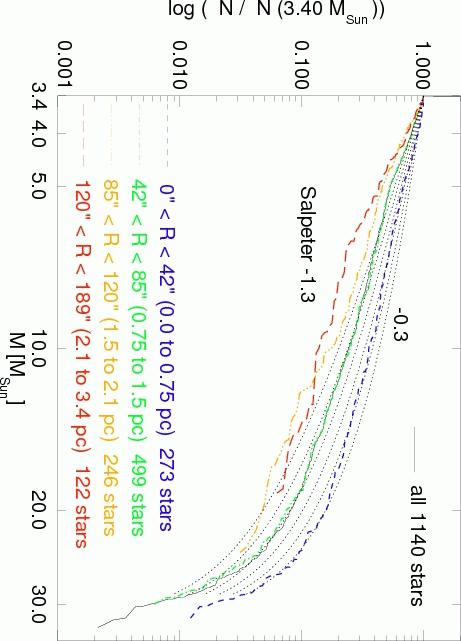}
\caption{Cumulative mass function for 4 annuli for stars with masses between 3.4 and 32\,M$_\odot$.}
\label{imf_cum}
    \end{figure}

\subsection{Mass function and number of cluster members}

Individual masses for cluster members are derived by comparison of their
location in the CMD with a 3.9\,Myr Geneva isochrone for stars
more massive than 6.0\,M$_\odot$, and a 3.2\,Myr PS99 isochrone for stars
with masses of 6.0\,M$_\odot$ and less. We note that in the evolutionary time scale defined by the tracks by Palla \& Stahler, a star with a mass of 3.5\,M$_\odot$ takes about 1.4\,Myr to cross from the birthline to the zero-age main 
sequence (ZAMS), while the location of the birthline for a star with 6.0\,M$_\odot$ coincides with the ZAMS.  No attempt was made to correct
for intrinsic infrared excess and/or extinction towards individual stars.
This is justified by the fact that a J--H vs.\ H--Ks colour-colour diagram
of the cluster members indicates that only a small fraction (less than 10\%)
have an intrinsic infrared excess. Stellar masses were assigned based
on the JHKs-brightness and -colours. For stars above 3.5\,M$_\odot$,
only those with J--Ks colours within $-0.3$ and $+0.4$\,mag of the isochrone
were considered for the mass function. As the pre-main sequence -- main-sequence
transition region marks an area with a broader spread in colour, for
stars below 3.5\,M$_\odot$, the J--Ks colour range was extended to allow
a variation of
$-0.3$ to $+0.6$\,mag around the colour indicated by the PS99 isochrone.

At an age of 3.9\,Myr, a star with an initial mass of 30\,M$_\odot$ has
lost about 3\% or 6.5\% of its initial mass for a metallicity of
Z = 0.02 or 0.04, respectively, according to the Geneva evolutionary
models. While this has a minute effect on the mass function, we assign
initial rather than present-day masses to each of the stars. 
The mass function obtained this way is then corrected for incompleteness,
and the slope of the mass function is computed for the range 3.4 to 
27\,M$_\odot$. 

Mass functions are traditionally visualised as histograms. Deriving the mass function slope from a histogram can be problematic, as the value of the fitted slope in general varies with bin size and locations (Ma\'{\i}z-Apell\'aniz \& \'Ubeda \cite{maiz05}).  Slopes derived from cumulative mass functions are much more robust (e.g., Stolte et al.\ \cite{stolte06}). In the following we discuss both 
standard histogram mass functions and cumulative mass functions.

Figure \ref{imf_iso} shows the mass function of the cluster
assuming solar metallicities for four annuli. The mass function slope $\Gamma$ has been fitted for the mass range 3.4 to 27\,M$_\odot$. 
With $\Gamma = -0.6$, the mass 
function is shallow in the inner $\approx$0.75\,pc  with respect to
a standard Salpeter mass function with $\Gamma = -1.35$, getting 
steeper for radial distances from the cluster centre between 0.75 and 1.5\,pc
with $\Gamma = -1.3$, $\Gamma = -1.6$ for radii between 1.5\,pc and 2.1\,pc,
and $\Gamma = -1.7$ for radii larger than 2.1\,pc. Because of the small
number of high mass stars, as well as uncertainties in the statistical
subtraction of field stars, the slopes in the two outermost annuli
have uncertainties of 5\% to 10\%, whereas the slopes in the inner annuli
have uncertainties smaller than 5\%.

Table \ref{imf_age} summarises the derived mass function slopes for masses between 3.4 and 27\,M$_\odot$. The lower limit of 3.4\,M$_\odot$ ensures that also in the innermost annulus only stars with masses above the 50\% completeness limit are included. The upper limit has been selected such that stars with  27\,M$_\odot$ are still below our photometric saturation limit even for an age of 4.9\,Myr.

The cumulative mass function for the same four annuli is shown in Fig.\ \ref{imf_cum}. It confirms that the slope of the mass function is getting steeper with increasing distance from the cluster centre. The mass function in the annulus with distances between 0.75 and 1.5 pc, i.e., the annulus centred on the half-mass radius (see section \ref{sec_hmr}) provides a good fit of the overall mass function of Wd\,1.
The MF slopes in the two outermost annuli (1.5 to 2.1\,pc and 2.1 to 3.4\,pc, 
respectively) is less well defined. As discussed already in the case of the
histograms, this is in part due to the relatively small
number of massive stars in the outermost annulus, and in part due to 
uncertainties in the subtraction of field stars towards lower masses. In
the following we thus assume an average slope of $\Gamma = - 1.6$ for the
two outermost annuli.

Assuming a single power-law initial mass function for the cluster in the mass range 0.5 to 120\,M$_\odot$ for each of the annuli, we can now estimate the total (initial) number of cluster stars. After correction for incompleteness, 1,385 stars with masses between 3.4 and 30 M$_\odot$
are present in the cluster. Below 0.5\,M$_\odot$ we assume a Kroupa IMF 
slope with $\Gamma = -0.3$. Table \ref{cluster_nstar} lists the total number of stars in the cluster. Including the $\approx$65 supernovae, which might 
already have exploded in Wd\,1 (see, e.g., Muno et al.\ \cite{muno06b}, and the
estimate below), the 
extrapolation indicates that we can expect a total of $\approx$100,000 stars 
with initial masses between 0.08 to 120\,M$_\odot$ in Wd\,1.

\subsection{Cluster mass}

\begin{table*}[htb]
\caption[]{Stars in the overlap between our sample based on NIR photometry and 
the optical spectroscopic studies  by Clark et al.\ (\cite{clark05}) and
Crowther et al.\ (\cite{crowther06}).} 
         \label{stars_spt_mass}
\begin{tabular}{crrcccccc}
W87 & RA(2000)& DEC(2000) & Ks & J-Ks & M$_{3.2{\rm Myr}}$& M$_{3.9{\rm Myr}}$ & M$_{4.9{\rm Myr}}$& Spt \\
\# & & & [mag]& [mag] & [M$_\odot$] & [M$_\odot$] & [M$_\odot$] & \\ \hline
60 & 16 47  4.12 &-45 51 52.3 & 9.13$\pm$0.01 &1.59$\pm$0.01& 36.7 & 32.0 &28.0& O9.5Ia - B0.5Ia$^a$  \\
29 & 16 47  4.41 &-45 50 40.0 & 9.17$\pm$0.03 &1.58$\pm$0.03& 36.4 & 31.7 &27.8& O9.5Ia - B0.5Ia \\
   & 16 47  6.54 &-45 50 39.2 & 9.25$\pm$0.05 &2.03$\pm$0.05&  &  && WR77s: WN6o  \\
15 & 16 47  6.63 &-45 50 29.8 & 9.66$\pm$0.01 &1.56$\pm$0.01& 30.6 & 27.6 &24.8& OB binary/blend \\
   & 16 47  7.65 &-45 52 36.0 & 9.33$\pm$0.02 &1.90$\pm$0.02&  &  && WR77sb(O): WN6o$^a$  \\
\end{tabular}
\begin{list}{}{}
\item[$^{\mathrm{a}}$] identified as a variable star (Bonanos \cite{bonanos07})
\end{list}

{\small The 1st column lists the identification according to Westerlund (\cite{west87}). Mass estimates (columns 7--9) are based on Geneva isochrones for ages of 3.2, 3.9 and 4.9\,Myr, respectively. Spectral types from  Clark et al.\ (OB-stars) and Crowther et al.\ (W-R stars) are
listed in column 10.}
\end{table*}

Table \ref{stars_spt_mass} lists bright stars close to the saturation limit of
Ks = 9.1\,mag in our sample, which also have spectral types determined by
 Clark et al.\ (\cite{clark05}). Mass estimates derived by comparison with Geneva isochrones show that these stars indeed have masses around 30\,M$_\odot$, as also determined by Clark et al.\ for the faint end of their sample.
This overlap between the two samples enables us to study the cluster mass over the entire range from our 50\% completeness limit up to the highest mass stars identified by Clark et al.\ (\cite{clark05}).

\begin{table*}[htb]
\caption[]{Integrated stellar mass for different annuli and stellar mass bins, and resulting total cluster mass including all stars detected on the SOFI frame
with individual mass estimates up to 30\,M$_\odot$.}
         \label{cluster_mass}
\begin{tabular}{cccc|c}
Radial Dist.\ & 0.8 -- 1.5 M$_\odot$ & 1.5 -- 3.5 M$_\odot$ & 3.5--30 M$_\odot$ & M$_{\rm tot}$  \\ 
 pc  & [M$_\odot$] & [M$_\odot$] &  [M$_\odot$] &  [M$_\odot$] \\ \hline
0 -- 0.75     & ? & $>$210   & 4135 &$>$4345  \\
0.75 -- 1.5   & $>$220 & 1090 & 4930 &$>$6240  \\
1.5 -- 2.1    & 455 & 780   & 1825 & 3060 \\
2.1 -- 3.3    & 260 & 340   & 540  & 1140 \\ \hline \hline
total mass    & $>$935 &$>$2420 & 11430 & $>$14785  \\  \hline
total mass$^a$ &        &      &       &$>$20800  \\ 
(0.8--120\,M$_\odot$)& &  &   \\ 
\end{tabular}
\begin{list}{}{}
\item[$^{\mathrm{a}}$] including the estimate of $\approx$6000\,M$_\odot$ in stellar mass
for stars more massive than 30\,M$_\odot$ (Clark et al.\ \cite{clark05}).
\end{list}

{\small
Mass bins below the 50\% completeness limit were not corrected for incompleteness. As a lower limit, only the total mass of stars actually detected in these bins is quoted, preceded by a ``$>$''. As the field of view is limited to an area of 4.3\,pc $\times$ 4.3\,pc,
approximately centred on the cluster, only stars located within 3.3\,pc of the cluster centre are considered.}
   \end{table*}

We determine the total stellar mass of the cluster in two slightly different 
ways. First, we add the mass of each star detected in the SofI data, correcting
this number by the completeness correction. The completeness correction 
was done on a star-by-star basis for each star detected, taking both the 
brightness and the radial distance from the cluster centre into account (i.e.\
for a 6\,M$_\odot$ star located in the very cluster centre, the correction 
factor applied is higher than for a 6\,M$_\odot$ star located near the outer 
edge of the innermost annulus). Table \ref{cluster_mass} summarises the stellar
mass in the cluster for three different annuli and stellar mass bins, only 
counting stars with masses between the respective
50\% completeness limits and 30\,M$_\odot$. Main sequence stars with masses between 3.5\,M$_\odot$ and 30\,M$_\odot$ amount to $\approx$11300\,M$_\odot$.
Including all low-mass stars with masses down to 0.8\,M$_\odot$, the total
stellar mass in this mass range is at least $\approx$14700\,M$_\odot$. If we
add to this the estimate from Clark et al.\ (\cite{clark05}) that stars more 
massive
than 30\,M$_\odot$ contribute another 6000\,M$_\odot$, the combined
stellar mass of Westerlund 1 amounts to at least $2 \times 10^4$\,M$_\odot$ (right column).
Higher resolution adaptive optics data, which we are going to discuss in a forthcoming paper, reveal that the cluster mass function extends well below 0.8\,M$_\odot$, adding even more stars and mass to the cluster.

\begin{table*}[htb]
\caption[]{Total number of stars in the cluster based on
extrapolation of the mass function slope, and scaling according to the observed number of stars with masses between 3.4 and 30\,M$_\odot$ for three annuli.} 
         \label{cluster_nstar}
\begin{tabular}{cc|cccccc|c}
Radial Dist.\ & $\Gamma_{\rm obs}$ & N$_{\rm obs}$ & N$_{\rm compl}$ &N$_{0.08-0.5}$ &N$_{0.5-30}$ &  N$_{30-50}$ & N$_{50-120}$ & N$_{\rm tot}$   \\
 pc  &        & (3.4--30\,M$_\odot$) & (3.4--30\,M$_\odot$) & (0.08--0.5\,M$_\odot$) & (0.5--30\,M$_\odot$) & (30--50\,M$_\odot$) &(50--120\,M$_\odot$) &(0.08--120\,M$_\odot$) \\ \hline
0 -- 0.75     & $-0.6$& 285 & 404 & 14800 & 1585 & 37 & 42 & 15460   \\
0.75 -- 1.5   & $-1.3$& 512 & 634 & 32700 & 8180 & 18 & 15 & 40910   \\
1.5 -- 3.5    & $-1.6$& 322 & 347 & 36700 & 7770 &  6 &  6 & 44480  \\ \hline \hline
sum           &       &1119 &1385 & 84200 & 17530& 61 & 63 & 101850  \\
\end{tabular}

{\small 
For the estimate in each annulus, a power-law IMF with the observed $\Gamma$ is assumed over the mass range from 0.5 to 120\,M$_\odot$, and $\Gamma = -0.3$ for masses between 0.08 and 0.5\,M$_\odot$.}
\end{table*}

\begin{table*}[htb]
\caption[]{Current and total initial stellar mass of the cluster based on
extrapolation of the mass function slope, and scaling according to the observed number of stars with masses between 3.4 and 30\,M$_\odot$ for three annuli.}
         \label{cluster_mass_imf}
\begin{tabular}{cc|cccc|c}
Radial Dist.\ & $\Gamma_{\rm obs}$ & M$_{0.08-0.5Mo}$ & M$_{0.5-30Mo}$ & Mi$_{30-50Mo}$ & M$_{50-120Mo}$ & M$_{\rm tot}$  \\
 pc  &        & [M$_\odot$] &[M$_\odot$] &[M$_\odot$] &[M$_\odot$] &[M$_\odot$] \\ \hline
0 -- 0.75     & $-0.6$& 3110 & 5460 & 1430 & 3210 & 13210  \\
0.75 -- 1.5   & $-1.3$& 6940 &12550 &  690 & 1080 & 21260  \\
1.5 -- 3.5    & $-1.6$& 7750 & 9400 &  275 &  430 & 17550 \\ \hline \hline
sum           &       &17800 &27410 & 2395 & 4720 & 52320 \\
\end{tabular}

{\small
For the estimate in each annulus, a power-law IMF with the observed $\Gamma$ is assumed over the mass range from 0.5 to 120\,M$_\odot$, and $\Gamma = -0.3$ for masses between 0.08 and 0.5\,M$_\odot$.}
\end{table*}

The second way to estimate the total stellar mass is based on the mass function
slopes, and the observed number of stars with masses between 3.4 and
30\,M$_\odot$, taking the completeness correction into account. By assuming a
single power law mass function in each of the cluster annuli for stellar masses
between 0.5\,M$_\odot$ (the peak of the Kroupa IMF) and 120\,M$_\odot$, 
and assuming $\Gamma = -0.3$ for masses between 0.08 and 
0.5\,M$_\odot$, we can extrapolate the total number
of stars and the total stellar mass. For the extrapolations we carried out 
Monte Carlo simulations assuming a randomly populated IMF. The simulations
yield a total number of stars of $\approx 100,000$ (Table 
\ref{cluster_nstar}), and a total initial stellar mass of 
$\approx 52,000$M$_\odot$ (Table \ref{cluster_mass_imf}) for the cluster,
with an uncertainty of about $\pm$15\%.
The Monte Carlo simulations also indicate that an uncertainty in the
mass function slope in the outermost annulus by $\pm0.1$ results in a change
of the total cluster mass by $\pm5$\%.
An additional uncertainty derives from the unknown binary properties of the 
cluster members. Both multiplicity fraction and
distribution of mass ratios need to be known in order to correct any derived
mass function and the total cluster mass for binary properties (see, e.g., the 
discussion in Stolte et al.\ \cite{stolte04,stolte06}). Our seeing limited
Sofi data with a resolution of 0.8$''$ (i.e. $\approx$2800\,A.U. at the distance of Wd\,1) do not provide strong constraints on the binary properties, so that
we currently cannot quantify the effect of unresolved binaries.

For comparison, C05 estimated a total mass of 100,000\,M$_\odot$ by assuming
that Wd\,1 currently houses 140 stars with initial masses $\ge$30\,M$_\odot$,
and then extrapolating down to stars with masses 0.08\,M$_\odot$ based on a
two-part Kroupa (\cite{kroupa02}) IMF with (in our notation) $\Gamma = -1.3$
for stars with masses $\ge$0.5\,M$_\odot$, and $\Gamma = -0.3$ for stars less
massive. 
Mengel \& Tacconi-Garman (\cite{mengel07}) derive a dynamical mass
 of $6.3^{+5.3}_{-3.7} \times 10^4$\,M$_\odot$ from the velocity dispersion
of four red supergiants in Wd\,1. Our estimate of a total initial cluster mass
of 52,000\,M$_\odot$ within $\approx$2.5\,pc of the cluster centre is in agreement with these studies. As there seem to be cluster members beyond 2.5\,pc
from the cluster centre, the actual cluster mass might be even higher.
This confirms that Wd\,1 is indeed one of the most massive 
starburst clusters in the
Milky Way, and one of the best local SSC analogues.

The extrapolation based on the observed slope and the single power law
mass function also indicates that the cluster initially had $\approx$65 
stars with masses between 50 and 120\,M$_\odot$. For a cluster 
age of $\approx$4\,Myr, these stars should already have turned into supernova. As
discussed by Muno et al.\ (\cite{muno06b}), this corresponds to a supernova
rate of one every $\approx$15,000\,yr. The main effect of the supernovae on the cluster
would have been the removal of any remnant gas tracing back to the formation of
the cluster as well as a loss of $\approx$4,700\,M$_\odot$ in stellar mass 
(corresponding to $\approx$10\% of the initial stellar mass in the cluster).
According to the IMF extrapolations, stars with masses between 30 and 
50\,M$_\odot$ should have initially contributed $\approx$2,400\,M$_\odot$
to the cluster mass. These stars, many of which are by now W-R stars,
should have lost a total of 1,000 to 1,500\,M$_\odot$. Thus at the high-mass
end, the cluster should have lost a total of 5,700 to 6,200\,M$_\odot$ in
stellar mass since its formation, resulting in a current cluster mass of
$\approx$45,000\,M$_\odot$.
Nevertheless, additional thus far undetected cluster members might be
located outside the field of view of the SofI data. Thus the actual mass of 
Wd\,1 could be even higher.

\subsection{Effect of age uncertainties and metallicity on mass function slope and cluster mass}

As discussed before, the uncertainty in the age determination of Wd\,1 traces 
back to the fact that we cannot a priori assume the Geneva and Palla \& Stahler
isochrones to be on the same absolute timescale. Based on the width and shape
of the pre-main sequence / main-sequence transition region, the best fitting 
PS99 isochrone can be determined with a small uncertainty in age. For stars
with masses above 6\,M$_\odot$, the selection of the best fitting Geneva
isochrone is not as straight forward.

Crowther et al.\ (\cite{crowther06}) discuss that Wd\,1 might have higher 
than solar metallicity as it is located at smaller galactocentric distances 
than the Sun and formed just recently. Based on the
analysis of the integrated spectrum of Wd\,1, however, Piatti et al.\ 
(\cite{piatti98}) conclude that it has close to solar metallicity.
In order to get an estimate on the effect of metallicity on the derived
mass function slope, we also applied a Geneva isochrone for an age of
3.9\,Myr and twice solar metallicity to the observations.

Table \ref{stars_spt_mass} exemplifies that the selection of different
Geneva isochrones with ages between $\approx$3 and 5\,Myr and solar 
metallicities as well as the selection of a twice solar metallicity isochrone
for an age of 3.9\,Myr results in a variation of the mass estimate by 
$\pm 15\%$ for the most massive stars in our sample. 
Towards lower masses this effect becomes less pronounced, and only the upper most bins of the mass function are affected. The estimate of the integrated cluster mass is even less subject to the uncertainties in the age estimate.  Table \ref{imf_age} summarises the different mass function slopes $\Gamma$ derived by applying Geneva isochrones with ages of 3.2, 3.9, and 4.9\,Myr, respectively, to the data.

From this we can conclude that the determination of the slope of the mass function as well as the cluster mass are quite robust, and relatively insensitive to the remaining uncertainties in the age and metallicity of Wd\,1.

\section{Structural and dynamical properties of Wd~1 \label{sec_hmr}}

\subsection{Cluster size and shape}

\begin{figure*}[htb]
   \centering
\hbox{
 \includegraphics[width=8.5cm,angle=0]{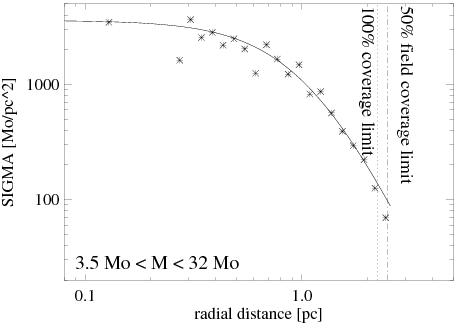}
 \includegraphics[width=8.5cm,angle=0]{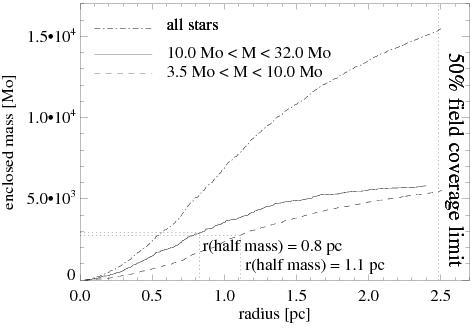}
}
\caption{Left: Radial surface mass density for stars with masses between 3.5 and 32\,M$_\odot$. The central surface density is $\ge 3.6 \times 10^3$M$_\odot$/pc$^2$. Overplotted is a function $ \propto (1+(r/a)^2)^{-2}$ with $a = 1.1$\,pc. Right: Enclosed stellar mass as a function of radial distance from the cluster centre. 
Mass segregation is evident by the increasing half-mass radius with decreasing 
stellar mass. For stars with masses between 10 and 32\,M$_\odot$ the half-mass radius is 0.8\,pc, whereas for stars with masses between 3.5 and 10\,M$_\odot$ 
the half-mass radius is 1.1\,pc.} 
\label{mass_rad}
\end{figure*}

The identification of several 1000 cluster members with masses between 0.8 and 32\,M$_\odot$ facilitates a study of the structural and dynamical properties of the cluster.

\begin{figure*}[htb]
   \centering
\hbox{
 \includegraphics[width=8.5cm,angle=0]{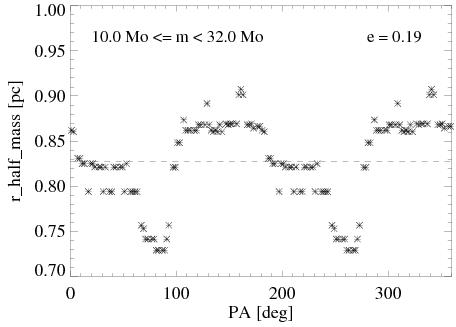}
 \includegraphics[width=8.5cm,angle=0]{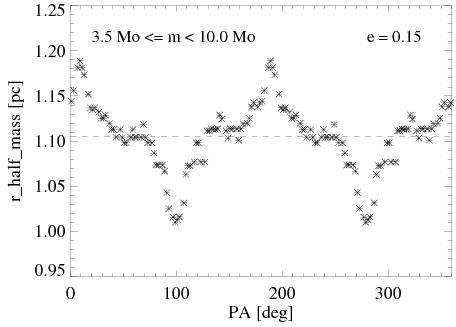}
}
\caption{The half mass radius as a function of position angle for stars with masses in the range 10 to 32\,M$_\odot$ (left) and masses in the range 3.5 to 10\,M$_\odot$ (right). The dashed line indicates the half mass radius averaged over all position angles. For both mass bins the cluster is clearly elongated, with a minor axis at PA $\approx$ 85$^\circ$ and an eccentricity = 0.19 in the high mass bin, and a minor axis at  PA $\approx$ 100$^\circ$ and an eccentricity of 0.15 in the low mass bin. The larger half-mass radius in the low-mass bin compared to the high mass bin indicates the presence of mass segregation. Note that we assumed the cluster half-mass radius to be point-symmetric with respect to the cluster centre.}
\label{cluster_shape}
    \end{figure*}

From the distance of 3.55\,kpc between the Sun and Wd\,1, a distance of 8.0\,kpc between the Sun and the GC, and the Galactic coordinates of Wd\,1, we compute its galactocentric distance to 4.8\,kpc. 
As already pointed out in the introduction, Wd\,1 is thus at a large enough distance from the GC that tidal effects leading to stripping of stars can be neglected.

Only counting stars with masses between 3.4 and 32\,M$_\odot$, the stellar surface density in the cluster core is at least $3.6 \times 10^3$\,M$_\odot$/pc$^2$.
The surface density profile follows a $ \propto (1+(r/a)^2)^{-2}$ law with $a = 1.1$\,pc (see Fig.\ \ref{mass_rad}, left). According to Elson et al.\ (\cite{elson87}, eq.\ 22) this corresponds to a core radius of 0.4\,pc. We do not attempt to fit a King profile, as the crowding and saturation of the brightest stars near the cluster centre prevents any accurate determination of the cluster's core radius based on the SofI data.

The half mass radius for all stars with masses in the range 3.4 to 32\,M$_\odot$, averaged over all position angles, is 1.0\,pc (Fig.\ \ref{mass_rad}, right).
Figure \ref{cluster_shape} gives evidence that the cluster is not spherical. Here we compute the half mass radius as a function of position angle, taking only stars located either in a quadrant $\pm 45^\circ$ relative to the position angle (PA) or located in the opposite quadrant (PA $\pm 180^\circ$) into account. 
The mass range of 3.0 to 32\,M$_\odot$ has been divided into two equal size bins in log mass. In both mass bins, the cluster is clearly elongated with its major axis along PA$\approx -10^\circ$ to $10^\circ$ and its minor axis along PA$\approx 80^\circ$ to $90^\circ$. 
The elongation is more pronounced for stars with masses in the range 10 to 32\,M$_\odot$ with an eccentricity of the cluster e = $1 - \frac{\rm minor axis}{\rm major axis}$ = 0.19 (Fig.\ \ref{cluster_shape}, left). The average half-mass-radius for stars in this mass bin is 0.8\,pc, which is in perfect agreement
with the half-light radius reported by Mengel \& Tacconi-Garman 
(\cite{mengel07}) based on another set of NTT/SofI observations of Wd\,1.
For stars with masses in the range 3.5 to 10\,M$_\odot$ the eccentricity of the cluster is ${\rm e} = 0.15$ and the average half mass radius is 1.1\,pc (Fig.\ \ref{cluster_shape}, right). 
The larger half mass radius in the lower mass bin indicates the presence of mass segregation.

An elongation was also reported by Muno et al.\ (\cite{muno06b}) in their 
analysis of the shape of the diffuse X-ray  emission from Wd\,1.
The diffuse X-ray emission shows an ellipticity of 0.75 (corresponding to an 
eccentricity of 0.25) with a major axis along 
PA = $13^\circ \pm 3^\circ$ and a characteristic radius (half-width at 
half-maximum) of 25$''$ (corresponding to 0.44\,pc for a distance of 
3.55\,kpc). Muno et al.\  (\cite{muno06b}) discuss possible
origins of the diffuse X-ray component. While the emission in the
core of Wd\,1 can be attributed to colliding winds of O and W-R stars with
initial masses $>$15\,M$_\odot$, less than 30\% of the extended diffuse 
emission can be explained by a spatially dispersed pre-main sequence star 
population with masses $\le$2\,M$_\odot$. A possible explanation for 
$\approx$70\% of the 
extended component of the diffuse emission are nonthermal particles
accelerated by colliding winds and supernova remnants.

The fact that both the amount of ellipticity and the direction of the major 
axis of the diffuse X-ray emission are in good agreement with the structural 
parameters derived from the NIR data for stars with masses between 10 and 
30\,M$_\odot$ supports the idea that colliding winds from the most massive
stars give rise to the diffuse X-ray emission in the core of Wd\,1.

In their study of the structure and dynamics of the ONC, Hillenbrand \&
Hartmann (\cite{hillenbrand98}) find an eccentricity of 0.30 for ONC. They
conclude that rotational flattening of ONC seems unlikely and that the
elongation might reflect the shape of the parental molecular cloud.

Given the still young age of Wd\,1 of 3 to 5\,Myr, its elongation, too,
should trace back to the formation
of the cluster. Possible explanations are formation out of an elongated 
parental molecular cloud, formation of the cluster out of two or multiple 
molecular sub-cores, out of a rotating molecular cloud, or a combination of these.  

\begin{table}[htb]
\caption[]{The decreasing half-mass radius with increasing stellar mass gives evidence for the presence of mass segregation in Wd\,1. The value in the mass bin 30--40\,M$_\odot$ is from C05.}
         \label{half_mass}
\begin{tabular}{c|ccc}
mass bin & 3--10\,M$_\odot$  & 10--30\,M$_\odot$ & 30-40\,M$_\odot$   \\ \hline
r$_{\rm hm}$  & 1.1\,pc & 0.8\,pc  & 0.44\,pc  (25$''$)  \\
\end{tabular}
\end{table}

\subsection{Cluster dynamics}

According to Binney \& Tremaine (\cite{binney87}), the velocity dispersion $\sigma$ of a relaxed stellar system in virial equilibrium is $$\sigma = \sqrt{0.4 \times {\rm G} \times {\rm M_{cl}} / {\rm r_{hm}}} $$
For a cluster mass of $\ge$ 45,000\,M$_\odot$ and a 2-D half-mass radius of ${\rm r_{hm}} = 1.0$\,pc (see Fig.\ \ref{mass_rad}, right), we derive a 1-D velocity dispersion $\sigma \ge 4.5{\rm km/s}$. 
This is in good agreement with the velocity dispersion of $5.0\pm 1.7$\,km/s
derived by Mengel \& Tacconi-Garman (\cite{mengel07}) based on a weighted
average for radial velocity measurements of four red supergiants in Wd\,1.
In  terms of proper motion, 4.5\,km/s corresponds to 0.25 mas/yr at a distance
of 3.55\,kpc. 
As shown by Stolte et al.\ (\cite{stolte07}) for the Arches cluster, multi-epoch adaptive optics observations at 8 to 10\,m class telescopes now achieve measurement accuracies better than 0.5\,mas/yr. Thus it should be possible to measure the velocity dispersion of a significant number of cluster stars using two epochs of high-angular resolution observations separated by a few years.
From this, any deviation from virial equilibrium due to the cluster's young age can be investigated. 
Rotation of the cluster as a whole might result in an anisotropy in the
observed velocity dispersion of stars. While in the direction parallel to
the rotation axis stars should just exhibit the internal velocity dispersion
of the cluster, in the direction perpendicular to the rotation axis stars
should on average show a larger velocity.

The crossing time for a star located at the half-mass radius and with a velocity equal to the velocity dispersion is t$_{\rm cross} = {\rm r_{hm}} / \sigma$ = $3 \times 10^5$\,yr. For a cluster age of 3.6\,Myr, this corresponds to $\approx$12 crossings since the formation of the cluster, 
indicating that the cluster already has experienced significant
dynamical relaxation and mass segregation.

Assuming a total number of cluster members N$_{\rm tot} = 10^5$ (Table \ref{cluster_nstar}), we can also compute the half-mass relaxation time (see, e.g., Spitzer \& Hart \cite{spitzer71}):
$$ {\rm t}_{\rm relax} = \frac{0.2 {\rm N_{\rm tot}}}{\ln (0.1 {\rm N_{\rm tot}})} \times \sqrt{0.4} \times {\rm t}_{\rm cross} = 4 \times 10^8\,yr $$ 

Thus even if the gaseous and stellar mass of the cluster would have been 
constant since its formation, it would not yet be dynamically relaxed, and 
further mass segregation would be expected. 
The expulsion of any remnant gas tracing back to the formation of the cluster 
by the multiple supernovae explosions, as well as the loss of 12\%
($\approx$6,000\,M$_\odot$) of the total initial stellar mass of the cluster
since its formation due to stellar winds and supernova explosions 
keep Wd\,1 out of virial equilibrium.

The effect of rotation on the dynamical evolution of a cluster was
discussed by Ernst et al.\ (\cite{ernst07}). They found that the
presence of rotation accelerates the dynamical evolution only in
systems with a single mass stellar component. In clusters with a realistic
mass function, this effect disappears as mass segregation becomes the primary
means to accelerate core collapse of the cluster. Thus we do not expect
that any rotation of Wd\,1 should have a strong effect on its dynamical
evolution.

As discussed, e.g., by Portegies Zwart et al.\ (\cite{portegies07}),
the presence and fraction of binary stars in a cluster can also have a strong 
effect on its dynamical evolution. Crowther et al.\ (\cite{crowther06}) discuss
evidence that at least between 1/3 to 2/3 of the W-R stars in Wd\,1
are members of binary systems. Bonanos (\cite{bonanos07}) identifies 
5 eclipsing binaries among the massive stars in Wd\,1. A systematic survey
for binaries and multiple systems in Wd\,1, in particular among the
intermediate to low mass stars, is still missing. Thus we currently
cannot quantify the binary fraction, and its effect on the dynamical
evolution of Wd\,1.


\section{Comparison to the NGC\,3603, Arches and 30 Doradus starburst clusters}

In Table \ref{sbc} we present a compilation and comparison of astrophysical 
quantities of the galactic young massive clusters Wd\,1 with ONC, NGC 3603YC, 
and Arches, as well as the \object{R136 cluster} in the \object{30\,Doradus region} in the \object{Large Magellanic Cloud}. 

\begin{table*}[htb]
\caption[]{Comparison of Westerlund\,1's astrophysical parameters with other young, massive clusters}
        \label{sbc}
\begin{tabular}{ccccccc}
 & Wd1 & ONC$^a$ & Wd2$^b$ & NGC 3603YC$^c$ & Arches$^d$ & R136 cluster$^e$  \\ \hline
Age [Myr]          &$3.6 \pm 0.7$ & 0.3--1.0 & 1--3 & 1--3  & 2--3 &2--3 \\
t$_{\rm relax}$ [Myr] & 400 &   2.1 &  &  & 2?  & \\
r$_{\rm core}$ [pc]  & 0.4 & 0.16--0.21 & 0.002--0.2 & ? & 0.034--0.30 \\
r$_{\rm hm}$ [pc]    &1.0 &0.8 &  &0.002--0.4 & 0.24  &1.1 \\
e                  &0.15--0.20 &0.30 & ? &? &? & ? \\
N$_{\rm tot}$      &$10^5$ & $>$2800 & &$>1680$ &$>$3450 & \\
M$_{\rm tot}$ [M$_\odot$] &$5 \times 10^4$ & $0.18 \times 10^4$ & $0.7 \times 10^4$ &$>0.7 \times 10^4$  &$>1.3 \times 10^4$ &$10 \times 10^4$ \\
(mass range in M$_\odot$) & 0.4--120 & 0.1--35 &0.08--120 & 0.1--120 &1--120 & 0.5--25 \\
\end{tabular}
\begin{list}{}{}
\item[$^{\mathrm{a}}$] Hillenbrand \& Hartmann (\cite{hillenbrand98}), and references therein.
\item[$^{\mathrm{b}}$] Ascenso et al.\ (\cite{ascenso07}), and references therein.
\item[$^{\mathrm{c}}$] Stolte et al.\ (\cite{stolte04,stolte06}), and references therein.
\item[$^{\mathrm{d}}$] Figer et al.\ (\cite{figer99}), Stolte et al.\ (\cite{stolte05}), and references therein.
\item[$^{\mathrm{e}}$] Andersen et al.\ (\cite{andersen07}), Brandl et al.\ (\cite{brandl96}), and references therein.
\end{list}
\end{table*}

All clusters have in common that they posses at least several O-type stars with masses in excess of 30\,M$_\odot$. 
The large number of evolved massive stars, including hypergiants and W-R stars, present in Wd\,1 gives evidence that it is the oldest, most evolved cluster in the sample.
Mass segregation seems to be a common feature among young clusters. As shown in
Fig.\ \ref{cluster_shape}, Wd\,1's half-mass radius increases with decreasing 
stellar mass. A similar trend has been observed by Brandl et al.\ (\cite{brandl96}) for the core radius of the R136 cluster. Compared to NGC\,3603YC and the Arches cluster, Wd\,1 has a rather large half-mass radius of $\approx$1.0\,pc, quite similar to the half-mass radii of the ONC and the R136 cluster.

Since dynamical mass estimates have not yet been derived for any of the clusters, and since the crowding and the presence of bright stars limits the ability to detect the low-mass stellar content of these clusters, mass estimates are in general lower limits. 
Wd\,1 with a total stellar mass between 21,000\,M$_\odot$ and 52,000\,M$_\odot$ is the most massive, and with a total stellar population of up to 100,000 stars also the most populous among the galactic young massive clusters presented in Table \ref{sbc}. Only the R136 cluster in the 30 Doradus region in the Large Magellanic Cloud, for which
Andersen et al.\ (\cite{andersen07}) derive a mass of 2.5 to 
3 $ \times 10^4$\,M$_\odot$ counting stars with masses down to 2.35\,M$_\odot$, is more massive.

\section{Summary and Outlook}

We have analysed near-infrared NTT/SofI observations of the starburst cluster
Westerlund 1, which is among the most massive young clusters in the Milky Way.
A comparison of colour-magnitude diagrams with theoretical main-sequence and
pre-main sequence evolutionary tracks yields improved extinction, distance
and age estimates of A$_{\rm Ks}$ =  1.13$\pm$0.03\,mag, d = 3.55$\pm$0.17\,kpc
(DM  = 12.75$\pm$0.10\,mag) and t = 3.6$\pm$0.7\,Myr, respectively. We derive
the slope of the stellar mass function for stars with masses between 3.4 and
27\,M$_\odot$. In an annulus with radii between 0.75 and 1.1\,pc from the
cluster centre, we get a slope of  $\Gamma = -1.3$, i.e.\ the Salpeter slope. 
Closer in, the mass function of Westerlund 1 is shallower with
$\Gamma = -0.6$, while at larger separations from the cluster it is getting steeper, reaching $\Gamma = -1.6$ for separations larger than 1.5\,pc. 
This is in good agreement with the change in mass function slope found in the 
starburst cluster NGC\,3603YC (Stolte et al.\ \cite{stolte06}).

Only considering stars with masses between 3.0 and 32\,M$_\odot$, we derive a half-mass radius of 1.0\,pc for the cluster. Wd\,1 exhibits clear deviations from spherical symmetry. The distribution of stars with masses between 10 and 32\,M$_\odot$ has an eccentricity of 0.20 with the major axis aligned roughly in the North-South direction. 
The distribution of stars with masses between 3.0 and 10\,M$_\odot$ is elongated in the same direction with an eccentricity of 0.15. The flattening of the cluster might be explained by rotation of the cluster along its minor axis.

By extrapolation of the observed mass function slopes for different annuli,
we derive an upper limit of the total initial stellar cluster mass of 
$\approx 52,000$M$_\odot$. Stellar evolution at the high mass end should have
reduced the total initial cluster mass by $\approx$6,000\,M$_\odot$ over the
past 4\,Myr.
By adding up the individual initial masses of stars directly detected in
the cluster, we derived a lower limit for the total present cluster mass of
m $\ge 2\times 10^4$\,M$_\odot$.

With a present-day mass of 20,000 to 45,000\,M$_\odot$ and an initial
stellar mass of $\approx$52,000\,M$_\odot$, Wd\,1 is the most massive
starburst cluster identified in the Milky Way to date, and about 10 times
as massive as the ONC, and 2 to 4 times as massive a NGC\,3603\,YC.
Additional cluster members located outside the field of view of the
SofI data might push the mass of Wd\,1 even higher.

In a following paper, we will report of the analysis of high-angular resolution
adaptive optics data, which trace the pre-main sequence population of the
cluster down to $\approx$0.2\,M$_\odot$. 

Further studies should aim at
determining proper motions for individual cluster members as well as
more radial velocity measurements. From this precise models of the cluster 
kinematics and
dynamics could be derived, which also might enable one to trace back the 
dynamical evolution of the cluster since its formation $\approx$4\,Myr ago.
Wd\,1, being the most massive, and one of the most nearby starburst clusters,
could turn into the prime example for studies on starburst cluster formation 
and evolution.

\begin{acknowledgements}
We would like to thank the referee S\/{\o}ren Larsen for the remarks, which helped to significantly improve the paper.
We are grateful to Paul Crowther, Andreas Ernst and Mike Muno for comments, which helped to improve the paper. WB acknowledges support by a Julian Schwinger fellowship of the University of California, Los Angeles. IN is supported by the Spanish Ministerio de Educaci\'{o}n y Ciencia
under grant AYA2005-00095.
  \end{acknowledgements}


\begin{thebibliography}{}

\bibitem[2007]{andersen07}
Andersen, M., Zinnecker, H., Moneti, A., Brandner, W., Brandl, B.\ et al., 2007, ApJ, subm.\
\bibitem[2007]{ascenso07}
Ascenso, J., Alves, J., Beletsky, Y., Lago, M.T.V.T.\ 2007, A\&A, 466, 137
\bibitem[2006]{bastian06}
Bastian, N., Goodwin, S.P., 2006, MNRAS, 369, L9
\bibitem[2007]{bonanos07}
Bonanos, A.Z., 2007, AJ, 133, 2696
\bibitem[1996]{brandl96}
Brandl, B., Sams, B.J., Bertoldi, F., et al., 1996, ApJ 466, 254
\bibitem[1999]{brandl}
Brandl, B., Brandner, W., Eisenhauer, F., Moffat, A. F. J., Palla, F., 
Zinnecker, H., 1999, A\&A, 352, L69
\bibitem[2000]{brandner00}
Brandner, W., Grebel, E., Chu, Y.-H., Dottori, H., Brandl, B.\ et al., 2000, AJ 119, 292
\bibitem[1987]{binney87}
Binney, J., Tremaine, S., 1987, Galactic Dynamics, Princeton Univ.\ Press, p.\ 214
\bibitem[2002]{cn}
Clark, J. S., Negueruela, I., 2002, A\&A, 396, L25
\bibitem[2005]{clark05}
Clark, J. S., Negueruela, I., Crowther, P. A., Goodwin, S. P., 2005, A\&A, 
434, 949
\bibitem[1996]{cotera96}
Cotera, A.S., Erickson, E.F., Colgan, S.W.J., et al., 1996, ApJ 461, 750
\bibitem[2006]{crowther06}
Crowther, P.A., Hadfield, L.J., Clark, J.S., Negueruela, I., Vacca, W.D., 2006, MNRAS 372, 1407
\bibitem[2007]{degrijs07}
de Grijs, R., Parmentier, G., 2007, ChJA\&A, 7, 155
\bibitem[2001]{devil01} 
Devillard, N.\ 2001, ESO C Library for an Image Processing Software Environment (eclipse), in ADASS X, ASP Conf.\ Ser.\ Vol.\ 238, eds.\ F.R.\ Harnden, F.A.\ Primini, H.E.\ Payne, p.\ 525
\bibitem[2005]{eisenhauer05}
Eisenhauer, F., Genzel, R., Alexander, T., Abuter, R., Paumard, T., et al.\ 2005, ApJ 628, 246
\bibitem[1987]{elson87}
Elson, R.A Fall, S.M., Freeman, K.C., 1987, ApJ 323, 54
\bibitem[2007]{ernst07}
Ernst, A., Glaschke, P., Fiestas, J., Just, A., Spurzem, R.\ 2007, MNRAS, 377, 465
\bibitem[1999]{figer99}
Figer, D.F., Kim, S.S, Morris, M., et al., 1999, ApJ 525, 750
\bibitem[2002]{figer02}
Figer, D.F., Najarro, F., Gilmore, D., Morris, M., Kim, S.S., et al.,
2002, ApJ 581, 258
\bibitem[2004]{figer04}
Figer, D.F., Rich, R.M., Kim, S.S., et al., 2004, ApJ 601, 319
\bibitem[2006]{hadfield06}
Hadfield, L.J., Crowther, P.A., 2006, MNRAS 368, 1822
\bibitem[1997]{hillenbrand97}
Hillenbrand, L.A., 1997, AJ 113, 1733
\bibitem[1998]{hillenbrand98}
Hillenbrand, L.A., Hartmann, L.W., 1998, ApJ 492, 540
\bibitem[2006]{kim06} 
Kim, S.S., Figer, D.F., Kudritzki, R.P., Najarro, F.\ 2006, ApJ 653, L113
\bibitem[2006]{klessen07} 
Klessen, R.S., Spaans, M., Jappsen, A.-K., 2007, MNRAS 374, L29
\bibitem[2007]{kothes07} 
Kothes, R., Dougherty, S.M., 2007, A\&A 468, 993
\bibitem[1960]{kron60} 
Kron, G.E., Mayall, N.U.\ 1960, AJ 65, 581
\bibitem[2002]{kroupa02a} 
Kroupa, P., 2002, Science 295, 82
\bibitem[2002]{kroupa02} 
Kroupa, P., Boily, C.M., 2002, MNRAS 336, 1188
\bibitem[2003]{lada}
Lada, C. J., Lada, E. A., 2003, ARA\&A, 41, 57
\bibitem[2006]{lamzin06}
Lamzin, S.A., 2006, AstL 32, 176
\bibitem[2004]{larsen}
Larsen, S. S., Richtler, T., 2004, A\&A, 427, 495
\bibitem[2006]{larson06}
Larson, R.B., 2006, RMXAA 26, 55
\bibitem[2001]{lejeune01}
Lejeune, T., Schaerer, D., 2001, A\&A, 366, 538
\bibitem[2005]{maiz05}
Ma\'{\i}z-Apell\'aniz, J., \'Ubeda, L., 2005, ApJ 629, 873
\bibitem[2007]{mayne07}
Mayne, N.J., Naylor, T., Littlefair, S.P., Saunders, E.S., Jeffries, R.D., 2007, MNRAS, 375, 1220
\bibitem[2005]{mccrady05}
McCrady, N., Graham, J.R., Vacca, W.D., 2005, ApJ 621, 278
\bibitem[2007]{mengel07}
Mengel, S., Tacconi-Garman, L.E., 2007, A\&A, 466, 151
\bibitem[1993]{meynet93}
Meynet, G., Mermilliod, J.-C., Maeder, A., 1993, A\&AS 98, 477
\bibitem[1993]{morris93}
Morris, M., 1993, ApJ 408, 496
\bibitem[2006a]{muno06a}
Muno, M.P., Clark, J.S., Crowther, P.A., et al.\ 2006a, ApJ 636, L41
\bibitem[2006b]{muno06b}
Muno, M.P., Law, C., Clark, J.S., et al.\ 2006b, ApJ 650, 203
\bibitem[2005]{nc}
Negueruela, I., Clark, J. S., 2005, A\&A, 436, 541
\bibitem[1999]{palla99} 
Palla, F., Stahler, S.W.\ 1999, ApJ 525, 772
\bibitem[2001]{patri01} 
Patriarchi, P., Morbidelli, L., Perinotto, M., Barbaro, G., 2001, A\&A 372, 644
\bibitem[1998]{piatti98} 
Piatti, A.E., Bica, E., Clari\'a, J.J., 1998, A\&AS 127, 423
\bibitem[2007]{portegies07} 
Portegies Zwart, S.F., McMillan, S.L.W., Makino, J., 2007, MNRAS 374, 95
\bibitem[1993]{reid93}
Reid, M.J., 1993, ARAA 31, 345
\bibitem[1985]{rieke85} 
Rieke, G.H., Lebofsky, M.J.\ 1985, ApJ 288, 618
\bibitem[1993]{rieke93} 
Rieke, G.H., Loken, K., Rieke, M.J., Tamblyn, P.\ 1993, ApJ 412, 99
\bibitem[2000]{siess00}
Siess L., Dufour E., Forestini M. 2000, A\&A, 358, 593
\bibitem[2006a]{skinner06a}
Skinner, S.L., Simmons, A.E., Zhekov, S.A.\ et al., 2006, ApJ 639, L35
\bibitem[2006b]{skinner06b}
Skinner, S.L., Perna, R., Zhekov, S.A., 2006, ApJ 653, 587
\bibitem[1971]{spitzer71}
Spitzer, L., Hart, M.H., 1971, ApJ 164, 399
\bibitem[1983]{stahler83}
Stahler, S.W., 1983, ApJ 274, 822
\bibitem[1987]{stetson87}
Stetson, P.B., 1987, PASP 99, 191
\bibitem[2002]{stolte02}
Stolte, A., Grebel, E., Brandner, W., Figer, D.,
2002, A\&A, 394, 459
\bibitem[2005]{stolte04}
Stolte, A., Brandner, W., Brandl, B., Zinnecker, H.,  Grebel, E.K., 2004, AJ 128, 765
\bibitem[2005]{stolte05}
Stolte, A., Brandner, W., Grebel, E. K., Lenzen, R., Lagrange, A.-M., 
2005, ApJ, 628, 113
\bibitem[2006]{stolte06}
Stolte, A., Brandner, W., Brandl, B., Zinnecker, H., 2006, AJ, 132, 253
\bibitem[2007]{stolte07}
Stolte, A., Ghez, A.M., Morris, M., Lu, J.R., Matthew, K., Brandner, W., 2007, ApJ, subm.
\bibitem[1961]{west61}
Westerlund, B. E., 1961, PASP, 73, 51
\bibitem[1987]{west87}
Westerlund, B. E., 1987, A\&AS, 70, 311
\bibitem[1995]{whitmore95}
Whitmore, B.C., Schweizer, F., 1995, AJ 109, 960
\bibitem[2001]{yi01}
Yi, S., Demarque, P., Kim, Y.-C., et al.\ 2001, ApJS 136, 417


\end{thebibliography}
\end{document}